\documentclass{article}
\usepackage[latin9]{inputenc}
\usepackage{geometry}
\usepackage{amsmath}
\usepackage{amssymb}
\usepackage{graphicx}
\usepackage{esint}

\makeatletter
\@ifundefined{date}{}{\date{}}

\newcommand{\qed}{\mbox{}\nolinebreak\hfill\rule{2mm}{2mm}\par\medbreak}
\newcommand{\Proof}{\noindent{\bf Proof. }}

\newcommand{\abs}[1]{\left\vert#1\right\vert}
\newcommand{\norm}[1]{\left\Vert#1\right\Vert}

 \newtheorem{theorem}{Theorem}
 \newtheorem{corollary}{Corollary}
 
 \newtheorem{remark}{Remark}

\@ifundefined{showcaptionsetup}{}{%
 \PassOptionsToPackage{caption=false}{subfig}}
\usepackage{subfig}
\makeatother

\begin{document}
\global\long\def\tanh#1{\,{\normalsize tanh}{\,#1}\,}
 \global\long\def\simbolo#1#2{#1\dotfill{#2}}
 \global\long\def\qed{\hbox to 0pt{}\hfill$\rlap{\ensuremath{\sqcap}}\sqcup$\medbreak}
 \global\long\def\theequation{\arabic{section}.\arabic{equation}}
 \global\long\def\thesection{\arabic{section}}

\title{Mathematical analysis of a model for moon-triggered clumping in Saturn's
rings}

\author{Pedro J. Torres%
\thanks{P.J. Torres is partially supported by project MTM2011-23652, Spain.
P. Madhusudhanan and L. W. Esposito were supported by the NASA Cassini
Mission, USA.%
}{\normalsize $^{*},$} Prasanna Madhusudhanan{\normalsize $^{\dagger},$}
Larry W. Esposito{\normalsize $^{\dagger}$}\\
\\
$^{*}$Departamento de Matem\'atica Aplicada, \\
 Universidad de Granada, Facultad de Ciencias, Granada, Spain. \\
\\
 $^{\dagger}$Laboratory of Atmospheric and Space Physics, \\
 University of Colorado, 392 UCB, Boulder, CO 80302, USA. \\
\\
 Email: ptorres@ugr.es, mprasanna@colorado.edu, larry.esposito@lasp.colorado.edu}
\maketitle
\begin{abstract}
\noindent Spacecraft observations of Saturn's rings show evidence
of an active aggregation-disaggregation process triggered by periodic
influences from the nearby moons. This leads to clumping and break-up
of the ring particles at time-scales of the order of a few hours.
A mathematical model has been developed to explain these dynamics
in the Saturn's F-ring and B-ring \cite{Es}, the implications of
which are in close agreement with the empirical results. In this paper,
we conduct a rigorous analysis of the proposed forced dynamical system
for a class of continuous, periodic and zero-mean forcing functions
that model the ring perturbations caused by the moon flybys. In specific,
we derive the existence of at least one periodic solution to the dynamic
system with the period equal to the forcing period of the moon. Further,
conditions for the uniqueness and stability of the solution and bounds
for the amplitudes of the periodic solution are derived.
\end{abstract}

\noindent \textbf{AMS2000 Subject Classification: 34C25, 85A99}.

\section{Introduction}

In \cite{Es}, Esposito et. al. have proposed a dynamical model to
explain the aggregation and disaggregation processes observed in Saturn's
F ring and the B ring outer edge due to the perturbation caused by
the Saturn's moon, Prometheus and Mimas, respectively. Their numerical
simulations demonstrate that the dynamical model explains certain
spacecraft measurements corresponding to these regions in the Saturn's
ring. Further, these numerical results strongly suggest the existence
of a limit cycle for realistic values of the parameters. In this paper,
we provide a rigorous mathematical proof of this fact as well as study
other features that are relevant from a dynamical point of view.

The model under consideration relates the mean aggregate mass $M$
and the velocity dispersion $v_{\mbox{\tiny rel}}^{2}$ of ring particles
at the above mentioned locations in the Saturn's ring, as follows

\begin{eqnarray}
\frac{dM}{dt} & = & \frac{M}{T_{\mbox{\tiny acc}}}-\left(\frac{v_{\mbox{\tiny rel}}^{2}}{v_{\mbox{\tiny th}}^{2}}\right)\frac{M}{T_{\mbox{\tiny coll}}}\nonumber \\
\frac{dv_{\mbox{\tiny rel}}^{2}}{dt} & = & -\frac{v_{\mbox{\tiny rel}}^{2}(1-\varepsilon^{2})}{T_{\mbox{\tiny coll}}}+\frac{M^{2}}{M_{0}^{2}}\frac{v_{\mbox{\tiny esc}}^{2}\left(M_{0}\right)}{T_{\mbox{\tiny stir}}}-f(t).\label{model}
\end{eqnarray}

In the first equation of (\ref{model}), the first term refers to
the coagulation of the ring particles leading to a growth in the mean
aggregate mass at an accretion rate $\frac{1}{T_{\mathrm{acc}}}$
($T_{\mbox{\tiny acc}}$ is the accretion period) and the second term
refers to the erosion (or break-up) of the larger ring particles due
to collisions with other ring particles at collisional rates $\frac{1}{T_{\mathrm{coll}}}$(
$T_{\mathrm{coll}}$ is the collisional period) when their velocity
dispersion exceed $v_{\mathrm{th}}$, the threshold velocity for sticking.

In the second equation of (\ref{model}), the first term refers to
the dissipative collisions between the ring particles leading to the
reduction in velocity dispersion with $\varepsilon$ as the normal
coefficient of restitution. The second term refers to the increase
in the velocity dispersion to the escape velocity from mass $M$ ($v_{\mathrm{esc}}\left(M_{0}\right)$
is the escape velocity from an aggregate of mass $M_{0}$) due to
the viscous stirring at a rate $\frac{1}{T_{\mathrm{stir}}}$ ($T_{\mathrm{stir}}$
is the stirring period), caused by the passage of the ring particles
of mass $M$. Finally, the last term refers to the periodic forcing
by the Saturn's moon causing disturbances in the velocity dispersion,
where $f(t)=\frac{2\pi v_{th}^{2}\beta}{T_{\mathrm{syn}}}\cos(\frac{2\pi t}{T_{\mathrm{syn}}})$
describes the moon perturbation, where $T_{\mathrm{syn}}$ is the
forcing period, $\beta$ is the forcing amplitude.

For convenience, we reduce the system to an equivalent dimensionless
system with the following substitutions: $x\equiv\frac{M}{M_{o}}$,
$y\equiv\frac{v_{\mathrm{rel}}^{2}}{v_{\mathrm{th}}^{2}}$ and $t\equiv\frac{t}{T_{\mathrm{orb}}}$,
where $T_{\mathrm{orb}}$ is the orbital period of the ring particles
around Saturn, and pass to a more mathematical notation by using $\frac{d}{dt}='$.
Then, the system in (\ref{model}) becomes
\begin{eqnarray}
x' & = & ax-bxy\nonumber \\
y' & = & -cy+dx^{2}-f(t)\label{model2}
\end{eqnarray}
where
\begin{equation}
a=\frac{T_{\mathrm{orb}}}{T_{\mbox{\tiny acc}}},\ b=\frac{T_{\mathrm{orb}}}{T_{\mbox{\tiny coll}}},\ c=\frac{T_{\mathrm{orb}}(1-\varepsilon^{2})}{T_{\mbox{\tiny coll}}},\ d=\frac{T_{\mathrm{orb}}v_{\mbox{\tiny esc}}^{2}}{T_{\mbox{\tiny stir}}v_{\mathrm{th}}^{2}}\label{param}
\end{equation}
are all positive parameters, $f\left(t\right)=A_{o}\cos\left(\omega t\right)$,
$A_{o}=\frac{2\pi\beta T_{\mathrm{orb}}}{T_{\mathrm{syn}}}$, $\omega=\frac{2\pi T_{\mathrm{orb}}}{T_{\mathrm{syn}}}=\frac{2\pi}{T}$,
and $T$ is the (normalized) forcing period.

The above dynamical system has two main differences when compared
to the predator-prey system. Firstly, in the second equation, we have
the $x^{2}$ term instead of the coupling $xy$. Secondly, the external
forcing is not standard in ecological models, since the typical forcing
is parametrical (see for instance the model studied in \cite{AO}).

Despite these differences, the aggregation-disaggregation model represented
by this system can be well explained by drawing an analogy with the
predator-prey system. Let the mean aggregate mass of the ring particles
correspond to the prey population, and the velocity dispersion correspond
to the predator population. The velocity dispersion 'feeds' off the
accelerations from the aggregates' gravity (in reference to the second
term of the second equation). As the velocity dispersion grows too
large, it limits the 'prey' as high velocity collisions lead to to
fragmentation of the ring particles (in reference to the second term
of the first equation). In the absence of the interaction between
the 'predator' and the 'prey', the 'prey' population (mean aggregate
mass) grows and the predator population decays (in reference to the
first terms of the two equations, respectively). In addition, the
predator population is periodically controlled in a certain deterministic
manner, as represented by the sinusoidal forcing function.

In the following section, we conduct a thorough qualitative analysis
of the above system and comment about the existence and nature of
its solution, conditions on uniqueness and asymptotic stability of
the solution and finally derive explicit bounds for the solutions.
Along with these results, we derive useful insights about the Saturn's
ring dynamics based on these results.

Further, the results in this paper hold for any general forcing functions
$f(t)$ as long as it is a continuous periodic function of minimal
period $T=\frac{2\pi}{\omega}$ with zero mean value, i.e., $\int_{0}^{T}f(t)dt=0$.
All the results will be presented for this general forcing function,
and the sinusoidal forcing function will be explicitly studied in
Section \ref{sec:example}.

\section{Existence of a periodic solution}

Numerical results in \cite{Es} strongly suggest the existence of
a limit cycle for realistic values of the parameters. In the following
theorem, we provide an analytical proof for the same. Further, the
result holds for all feasible values of the system parameters.

\begin{theorem}\label{th1} System $(\ref{model2})$ has at least
one $T$-periodic solution $(x,y)$ with $x(t)>0$ for all $t$. \end{theorem}

\Proof By introducing the change $z=\ln x$, system $(\ref{model2})$
is transformed into
\begin{equation}
\begin{array}{lcl}
z' & = & a-by\\
\\
y' & = & -cy+d\exp(2z)-f(t).
\end{array}\label{model3}
\end{equation}

The theorem is proved by showing that the above system is equivalent
to a second order differential equation of the Duffing type, and then
invoking the Landesman-Lazer conditions for the existence of a periodic
solution. See Appendix \ref{sub:proofth1} for the complete proof.
\qed

The above result leads to a physical interpretation that the periodic
forcing by the moon gives a limit cycle in the mean aggregate mass
and velocity dispersion with the same period as the forcing function.

It is interesting to remark that the latter result can be extended
to a family of forcing terms with more complicated recurrence including
quasi periodic forcing. The main result in \cite{Ah} implies the
existence of a quasi periodic solution and the boundedness of all
solutions in the future.

\section{Uniqueness and asymptotic stability}

In this section, we derive conditions for the existence of a unique
periodic solution and comment about the stability of the solution.

\begin{theorem}\label{stab1} When
\begin{equation}
acT<2+\frac{c^{2}}{2},\label{cond1}
\end{equation}
system $(\ref{model2})$ has a unique $T$-periodic solution, which
is asymptotically stable. \end{theorem}

\Proof See Appendix \ref{sec:proofStab1}.\textbf{\qed}

\begin{remark} Observe that for $f\equiv0$, the unique equilibrium
of system $(\ref{model2})$ is hyperbolic, hence robust to small perturbations.
Therefore, if $A_{0}$ is small enough, $(\ref{model2})$ presents
an asymptotically stable $T$-periodic orbit. The remarkable feature
of the latter result is that stability is proved on a certain region
of the involved parameters independently of the size of the external
force $f$.\end{remark}

As we will see in Section \ref{sec:example}, Theorem \ref{stab1}
is applicable to the Saturn's F-ring, in fact for the $F$-ring $ac=16\tau^{2}(1-\varepsilon^{2})\simeq0.104$
so condition $(\ref{cond1})$ is satisfied. The condition $(\ref{cond1})$
is also satisfied for the Saturn's B-ring outer edge when the optical
depth is $\tau=.5$. In the following theorem, a condition for the
existence and stability of the solution is derived that depends on
the nature of the forcing function via its supremum norm, denoted
by $\norm{f}_{\infty}$.

\begin{theorem}\label{stab2} Under the assumption
\begin{equation}
2ac+2b\norm{f}_{\infty}<\frac{c^{2}}{4}+\frac{\pi^{2}}{T^{2}},\label{cond2}
\end{equation}
system $(\ref{model2})$ has at least one asymptotically stable $T$-periodic
solution. \end{theorem}

\Proof The above result is obtained by using the technique of lower
and upper solutions \cite{NO}. See Appendix \ref{sec:proofStab2}
for the complete proof.\qed

In the context of the Saturn's ring dynamics, Theorem \ref{stab1}
and Theorem \ref{stab2} show that the forced ring system has a limit
cycle behavior around the fixed point and when the external forcing
satisfies the above condition, it drives a stable limit cycle. Further,
the above theorems provide a theoretical justification for the concentric
limit cycles that were observed through numerical simulations in \cite[Fig. 16 and Fig. 17]{Es}.

\section{Explicit bounds}

Theorem \ref{th1} provides a general existence result, but more concrete
quantitative information about the location of the periodic orbits
is desirable. In this section, we derive explicit estimates for the
periodic solutions. In the next result, $f^{+}=\max\{f,0\}$, $f^{-}=\max\{-f,0\}$
denotes the positive and negative part respectively of a given function
$f$, and $\norm{f}_{p}$ denotes the usual $L^{p}$-norm.

\begin{theorem}\label{th2} Let $(x,y)$ with $x(t)>0$ for all $t$
a $T$-periodic solution of system $(\ref{model2})$. Then, the following
bounds hold
\begin{equation}
\begin{array}{rcl}
\sqrt{\frac{ac}{bd}}\exp\left(-\frac{\sqrt{T}}{2}\frac{b}{c}\norm{f}_{2}\right) & \leq x(t)\leq & \sqrt{\frac{ac}{bd}}\exp\left(\frac{\sqrt{T}}{2}\frac{b}{c}\norm{f}_{2}\right)\\
\frac{a}{b}(1-cT)-\norm{f^{+}}_{1} & \leq y(t)\leq & \frac{a}{b}(1+cT)+\norm{f^{-}}_{1}
\end{array}\label{bounds}
\end{equation}
for all $t$. \end{theorem}

\Proof The bounds derived above are obtained by working with the
second order differential equation of Duffing type (\ref{duff}) that
is equivalent to the system in (\ref{model2}). See Appendix \ref{sec:proofth2}
for the complete proof.\qed

Since the lower bound for $x\left(t\right)$ is non-negative, the
$T$-periodic solution $x\left(t\right)$ is also non-negative for
all $t$. The above argument acts as an alternate proof for the later
part of the statement in Theorem \ref{th1}. Further, the bounds derived
for the $T$-periodic solution for $y\left(t\right)$ may not necessarily
be tight. Especially, for small forcing amplitudes, the $T$-periodic
solutions are expected to be close to the stable feasible fixed point,
but the upper and lower bounds deviate from the stable fixed point
by a factor of $cT$ about the fixed point. As a result, we derive
tighter bounds for $y\left(t\right)$ in the following theorem.

\begin{theorem}\label{th3}Alternatively, $y$ satisfies the bounds:

\begin{eqnarray}
 &  & y\left(t\right)\ge\max\left(\frac{ab^{-1}cT}{\left(\mathrm{e}^{cT}-1\right)},\frac{a\exp\left(-\frac{b\sqrt{T}\left\Vert f\right\Vert _{2}}{c}\right)}{b}\right)-\min\left(v_{max},\sqrt{\frac{\left\Vert f\right\Vert _{2}^{2}\left(\mathrm{e}^{cT}+1\right)}{2c\left(\mathrm{e}^{cT}-1\right)}},\frac{\left\Vert f^{+}\right\Vert _{\infty}}{c},\frac{\left\Vert f^{+}\right\Vert _{1}}{\left(1-\mathrm{e}^{-cT}\right)}\right),\label{eq:ylb}\\
 &  & y\left(t\right)\le\min\left(\frac{ab^{-1}cT}{\left(1-\mathrm{e}^{-cT}\right)},\frac{a\exp\left(\frac{b\sqrt{T}\left\Vert f\right\Vert _{2}}{c}\right)}{b}\right)-\max\left(v_{min},\frac{-\left\Vert f^{-}\right\Vert _{\infty}}{c},\frac{-\left\Vert f^{-}\right\Vert _{1}}{\left(1-\mathrm{e}^{-cT}\right)}\right),\label{eq:yub}
\end{eqnarray}
where $v_{max}=\underset{t}{\max}\int_{s=t}^{t+T}G\left(t,s\right)f\left(s\right)ds$,
$v_{min}=\underset{t}{\min}\int_{s=t}^{t+T}G\left(t,s\right)f\left(s\right)ds$,
and $G(t,s)=\frac{\exp(cs-ct)}{\exp(cT)-1}$ is the Green's function.

\end{theorem}

\Proof The alternative bounds given by $(\ref{eq:ylb})$ and $\left(\ref{eq:yub}\right)$
are found working directly with the second equation of system $(\ref{model3})$.
Given a $T$-periodic solution $h(t)$, it is known that the unique
$T$-periodic solution of the first-order linear equation $y'+cy=h(t)$
is
\[
y=\int_{t}^{t+T}G(t,s)h(s)ds.
\]
Then, from the second equation of system $(\ref{model3})$, we have
\[
y=\int_{t}^{t+T}G(t,s)\left[d\exp(2z)-f(s)\right]ds.
\]
Tight upper and lower bounds are derived for each of the two terms
in the above integral considered separately. By appropriately combining
these bounds, we obtain $(\ref{eq:ylb})$ and $\left(\ref{eq:yub}\right)$.
See Appendix \ref{sec:proofth3} for the complete proof. \qed

The bounds in $(\ref{eq:ylb})$ and $\left(\ref{eq:yub}\right)$ are
tight for small forcing amplitudes. Combining these with the second
inequality in (\ref{bounds}), the bounds are tight for all forcing
amplitudes.

\begin{remark}

Since $y$ represents a squared relative velocity, a useful lower
bound for $y$ should be positive. In this sense, the second bound
given by $(\ref{eq:ylb})$ has the advantage that the lower bound
is always positive if the external force $f$ is not too large. In
the model case $f(t)=A_{0}\cos(\omega t)$, it is easy to compute
$\norm{f^{+}}_{1}=\norm{f^{-}}_{1}=2A_{0}/\omega$, $\norm{f}_{2}=A_{0}\sqrt{\frac{\pi}{\omega}}$,
and the bounds are direct.

\end{remark}

Having completed the mathematical analysis of the system (\ref{model2}),
in the following section, we consider examples with realistic values
of the parameters of the system representing different regions of
the Saturn's ring and derive useful insights pertinent to the ring
dynamics.

\section{\label{sec:example}Insights on Saturn's ring dynamics}

We begin with some corollaries of the theorems presented in the previous
sections for the Saturn's ring system.

\begin{remark}

\label{rem:fixedPoint}The non-trivial, feasible stable fixed point
for the unforced dynamic system is $\left(M,v_{\mathrm{rel}}^{2}\right)=\left(\sqrt{\frac{T_{\mathrm{stir}}\left(1-\varepsilon^{2}\right)}{v_{\mathrm{esc}}^{2}\left(M_{o}\right)T_{\mathrm{acc}}}},\ \frac{T_{\mathrm{coll}}}{T_{\mathrm{acc}}}\right)$.
When externally forced by periodic perturbations from the Saturn's
moons, the mean aggregate mass and the velocity dispersion demonstrates
periodic aggregation-disaggregation cycle with the same period as
that of the moon's forcing.

\end{remark}

\begin{corollary} (of Theorem \ref{stab1}) There exists a unique,
asymptotically stable limit-cycle for the ($M$,$v_{\mathrm{rel}}^{2}$)
system if the forcing period satisfies the condition
\begin{eqnarray}
T & < & \frac{1+4\tau^{2}\left(1-\varepsilon^{2}\right)^{2}}{8\tau^{2}\left(1-\varepsilon^{2}\right)},\label{eq:fsynLowerBound}
\end{eqnarray}
where the right-hand-side depends on the mean optical depth and the
normal coefficient of restitution corresponding to the specific location
in the Saturn's ring system.

\end{corollary}

The realistic parameters corresponding to the Saturn's ring system
are $\varepsilon=0.6$, $T_{\mathrm{syn}}=T_{\mathrm{orb}}=T=1$.
The optical depth is $\tau=0.1$ for the F-ring and $\tau=1.5,\ 1,\ 0.5$
for the B-ring outer edge. Hence, our result is applicable to the
F-ring and also to the B-ring outer edge when $\tau=0.5$.

Now, we consider a systematic study of each region in the Saturn's
rings, and start with the dynamics of the B-ring outer edge.

\begin{corollary}

For a sinusoidal forcing function, $f(t)=\frac{2\pi\beta}{T}\cos\left(\frac{2\pi t}{T}\right)$,
the upper and lower bounds of the $T$-periodic solution for the mass
of the aggregate is
\begin{eqnarray}
 &  & \mathrm{e}^{\frac{-\pi\beta}{\sqrt{2}\left(1-\varepsilon^{2}\right)}}\le\frac{M\left(t\right)}{2\sqrt{4\tau T_{\mathrm{stir}}\left(1-\varepsilon^{2}\right)}}\le\mathrm{e}^{\frac{\pi\beta}{\sqrt{2}\left(1-\varepsilon^{2}\right)}}.\label{eq:b1}
\end{eqnarray}
For $v_{\mathrm{rel}}^{2}\left(t\right)$, the bounds are obtained
from Theorems \ref{th2} and \ref{th3} with $a=b=\frac{c}{1-\varepsilon^{2}}=4\tau$,
$v_{\mathrm{esc}}\left(M_{0}\right)=0.5$m/sec, $v_{\mathrm{th}}=1$m/sec,
$v_{max}=-v_{min}=\frac{2\pi\beta}{c\sqrt{c^{2}T^{2}+4\pi^{2}}}$,
$\left\Vert f\right\Vert _{2}=\sqrt{\mbox{\ensuremath{\int}}_{t=0}^{T}\left|f\left(t\right)\right|^{2}dt}=\pi\beta\sqrt{\frac{2}{T}}$,
$\left\Vert f^{+}\right\Vert _{1}=\mbox{\ensuremath{\int}}_{t=0}^{T}\left|f^{+}\left(t\right)\right|dt=2\beta$,
$,\left\Vert f^{+}\right\Vert _{1}=\mbox{\ensuremath{\int}}_{t=0}^{T}\left|f^{-}\left(t\right)\right|dt=2\beta$,
and $\left\Vert f\right\Vert _{\infty}=\frac{2\pi\beta}{T}$.

\end{corollary}

\begin{figure}
\begin{centering}
\subfloat[\label{fig:fstirFig1} ]{\begin{raggedright} \includegraphics[scale=0.6]{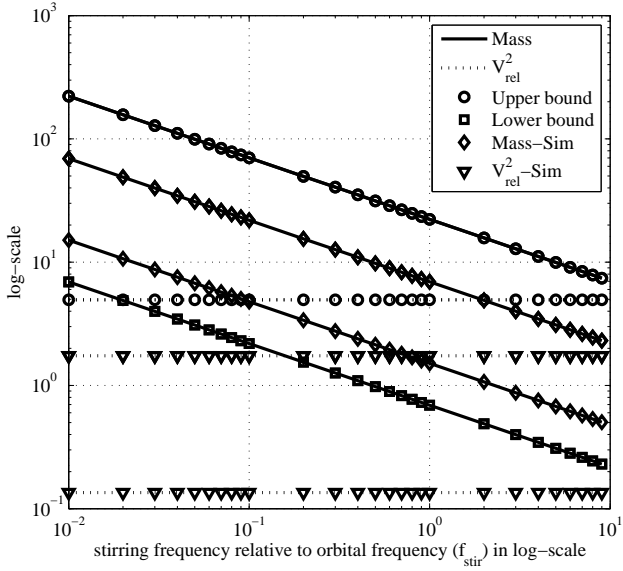}

\end{raggedright}

}\subfloat[\label{fig:fstirFig2}]{\begin{raggedright} \includegraphics[scale=0.6]{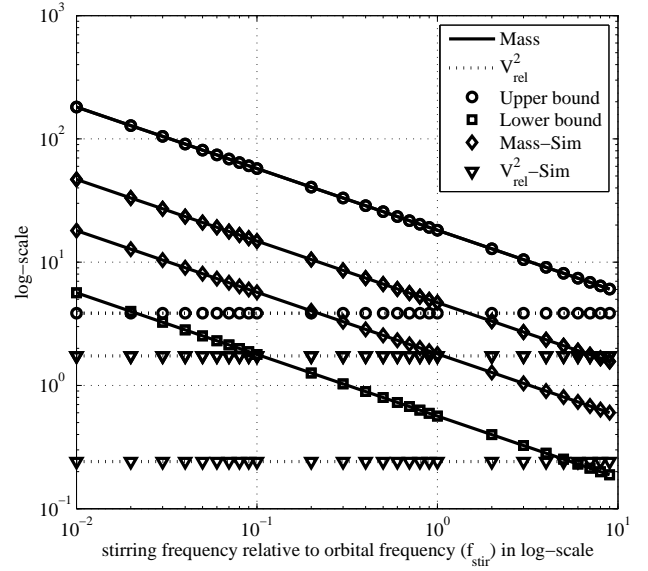}

\end{raggedright}

}
\par\end{centering}

\begin{centering}
\subfloat[\label{fig:fstirFig3}]{\begin{raggedright} \includegraphics[scale=0.6]{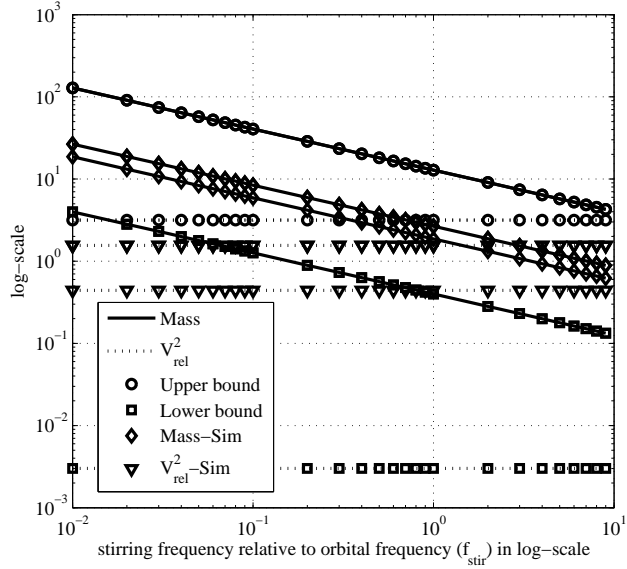}

\end{raggedright}

}
\par\end{centering}


\caption{\textbf{Derived bounds compared to numerical simulations for varying
$f_{\mathrm{stir}}$: }The system parameters corresponds to the B-ring
outer edge where the normal coefficient of restitution $\varepsilon=0.6$,
the forcing period of the moon is equal to the orbital period of the
ring particles about Saturn ($T=1$) and the forcing amplitude $\beta=0.5$.
(a) The mean optical depth is $\tau=1.5$, (b) the mean optical depth
is $\tau=1$ and (c) the mean optical depth is $\tau=0.5$. }

\end{figure}

We begin with studying the impact of varying the stirring rate $f_{\mathrm{stir}}=\frac{1}{T_{\mathrm{stir}}}$
on the amplitudes of the periodic solutions of $\left(M\left(t\right),v_{\mathrm{rel}}^{2}\left(t\right)\right)$,
and Figures \ref{fig:fstirFig1} - \ref{fig:fstirFig3} plot the maxima
and minima of the $T$-periodic solution of the $\left(M,v_{\mathrm{rel}}^{2}\right)$
system obtained using numerical simulations against the derived upper
and lower bounds for the $T$-periodic solution of the $\left(M,v_{\mathrm{rel}}^{2}\right)$
system from Theorem \ref{th2} and Theorem \ref{th3}. The system
parameters corresponding to the Saturn's ring system are $\varepsilon=0.6$,
$T_{\mathrm{syn}}=T_{\mathrm{orb}}$ $\left(T=1\right)$, $\beta=0.5$
and $\tau=1.5,\ 1,$ and $0.5$ for Figure \ref{fig:fstirFig1}, Figure
\ref{fig:fstirFig2} and Figure \ref{fig:fstirFig3}, respectively.
The Saturn's B-ring including the Janus 2:1 density wave, and the
B-ring outer edge have 1.5 and 1 as the typical values for the optical
depth. Further, $\tau=0.5$ is the typical optical depth value for
the Saturn's A ring, Mimas 5:3 density wave and the Janus 6:5 density
wave.

From (\ref{eq:b1}), it is clear that the amplitude upper and lower
bounds for the periodic solutions of $M\left(t\right)$ vary log-linearly
with $f_{stir},$ and hence the corresponding curves in the plot are
straight lines. It is interesting to note that the actual values for
the maximum and minimum amplitudes (obtained through simulations)
also appear to have a log-linear relationship with $f_{stir}$ with
the same slope as that of the upper and lower bounds. As a result,
an exponential increase in $f_{stir}$ causes an exponential decay
in the mean aggregate mass of the ring particles. This effect can
be understood as follows. Large $f_{stir}$ causes the rate of change
of $y\left(t\right)$ to be positive, and once $v_{\mathrm{rel}}^{2}\left(t\right)>\frac{T_{coll}}{T_{acc}}$
(same as $y>\frac{a}{b}$ in (\ref{model3})), this causes a negative
slope for $M\left(t\right),$ and essentially causing the exponential
decay of $M\left(t\right).$ Further, the parameter $f_{stir}$ has
a negligible effect on the amplitudes of $v_{\mathrm{rel}}^{2}\left(t\right),$
as is evident from the negligible slope in the figure above. Notice
that both the axes are in log-scale and the plots for the lower-bound
of $v_{\mathrm{rel}}^{2}\left(t\right)$ does not appear in the plot
because it takes negative values for all values of $f_{stir}.$

We see that the effect of changing $\tau$ only changes the magnitudes
of maxima, minima, and the bounds on $M\left(t\right)$ and $v_{\mathrm{rel}}^{2}\left(t\right).$
Otherwise, $M\left(t\right)$ has a log-linear relationship with $f_{stir},$
with the same slope for all values of $\tau$. Comparing Figures \ref{fig:fstirFig1}
- \ref{fig:fstirFig3}, the magnitudes of maxima $M\left(t\right)$
increases with increase in $\tau,$ and the minima decreases with
increase in $\tau.$ Further, the magnitude of the maxima of $v_{\mathrm{rel}}^{2}\left(t\right)$
remains unchanged with increase in $\tau$ and the magnitude of the
minima decreases with increase in $\tau.$ The physical interpretation
is that larger optical depth leads to a larger mean aggregate mass,
but not large velocity dispersion. This is because the velocity dispersion
is pegged near the fixed point value given by the strong erosion for
$v_{\mathrm{rel}}^{2}\left(t\right)>v_{\mathrm{th}}^{2}$, which reverses
the growth of $M\left(t\right)$. For the same reason, increasing
$f_{\mathrm{stir}}$ reduces the upper value for $M\left(t\right)$.

\begin{figure}
\begin{centering}
\subfloat[\label{fig:betaFig1}]{\begin{centering}
\includegraphics[scale=0.6]{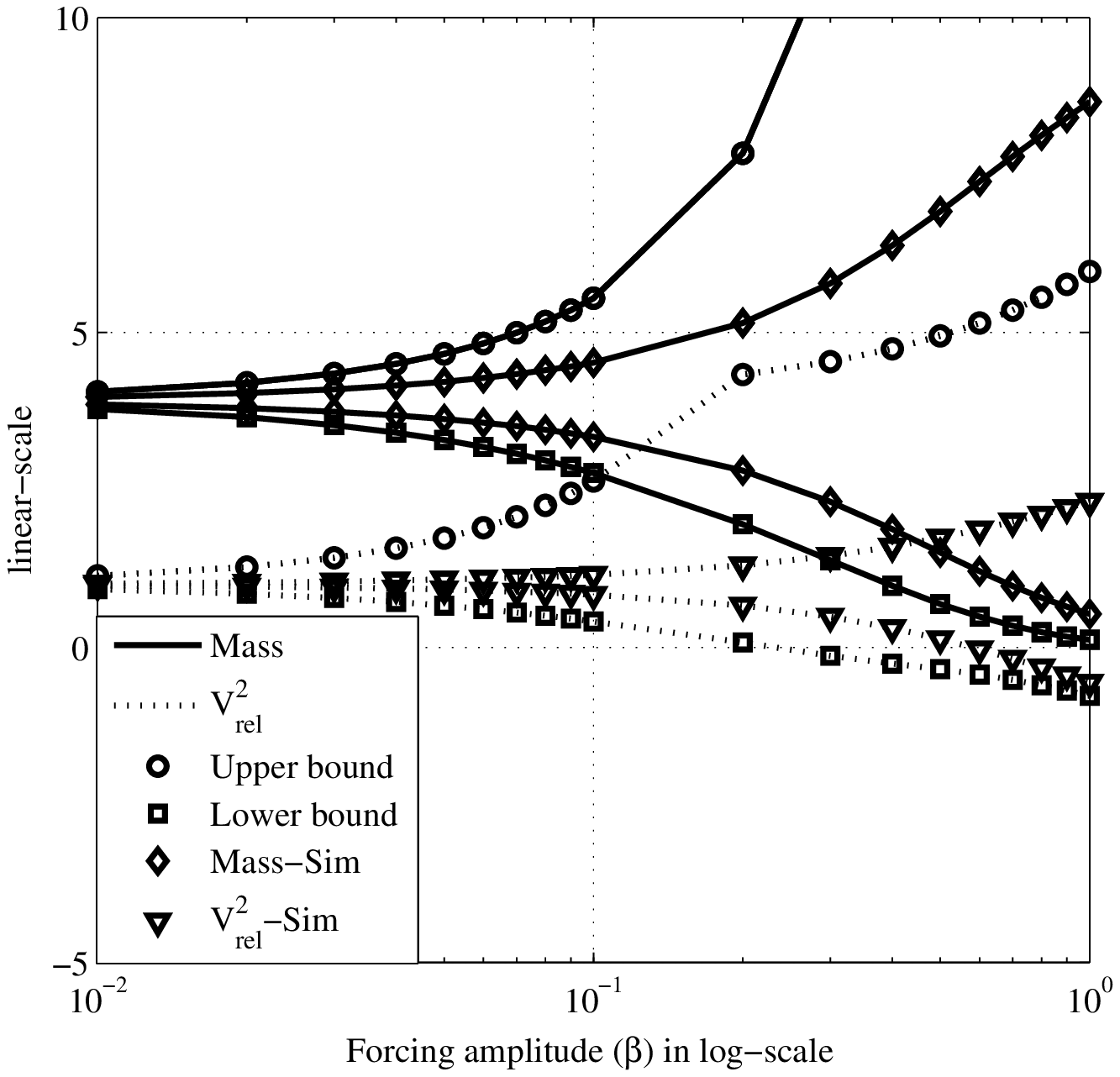}
\par\end{centering}

}\subfloat[\label{fig:betaFig2}]{\begin{centering}
\includegraphics[scale=0.6]{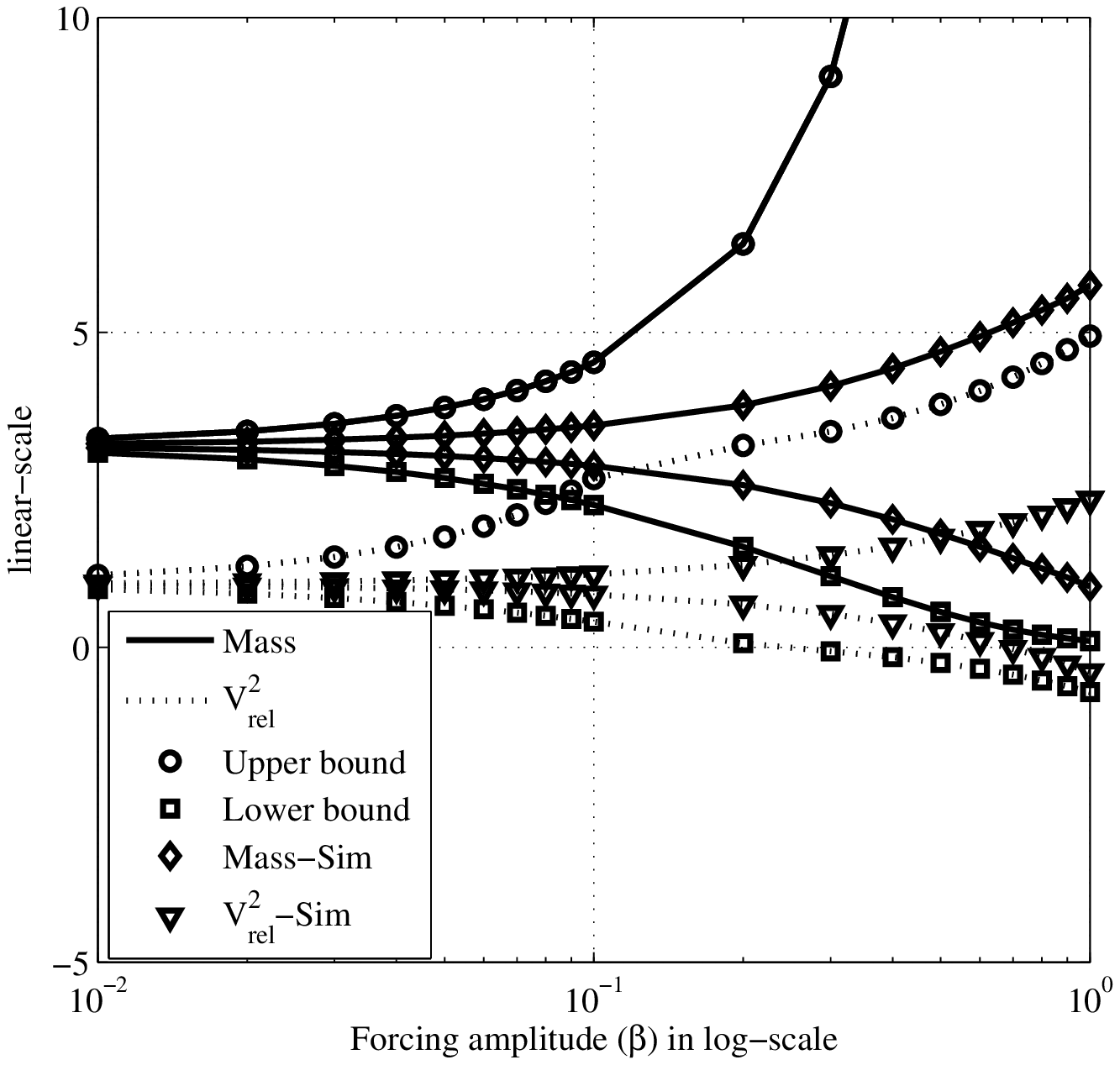}
\par\end{centering}

}
\par\end{centering}

\begin{centering}
\subfloat[\label{fig:betaFig3}]{\begin{centering}
\includegraphics[scale=0.6]{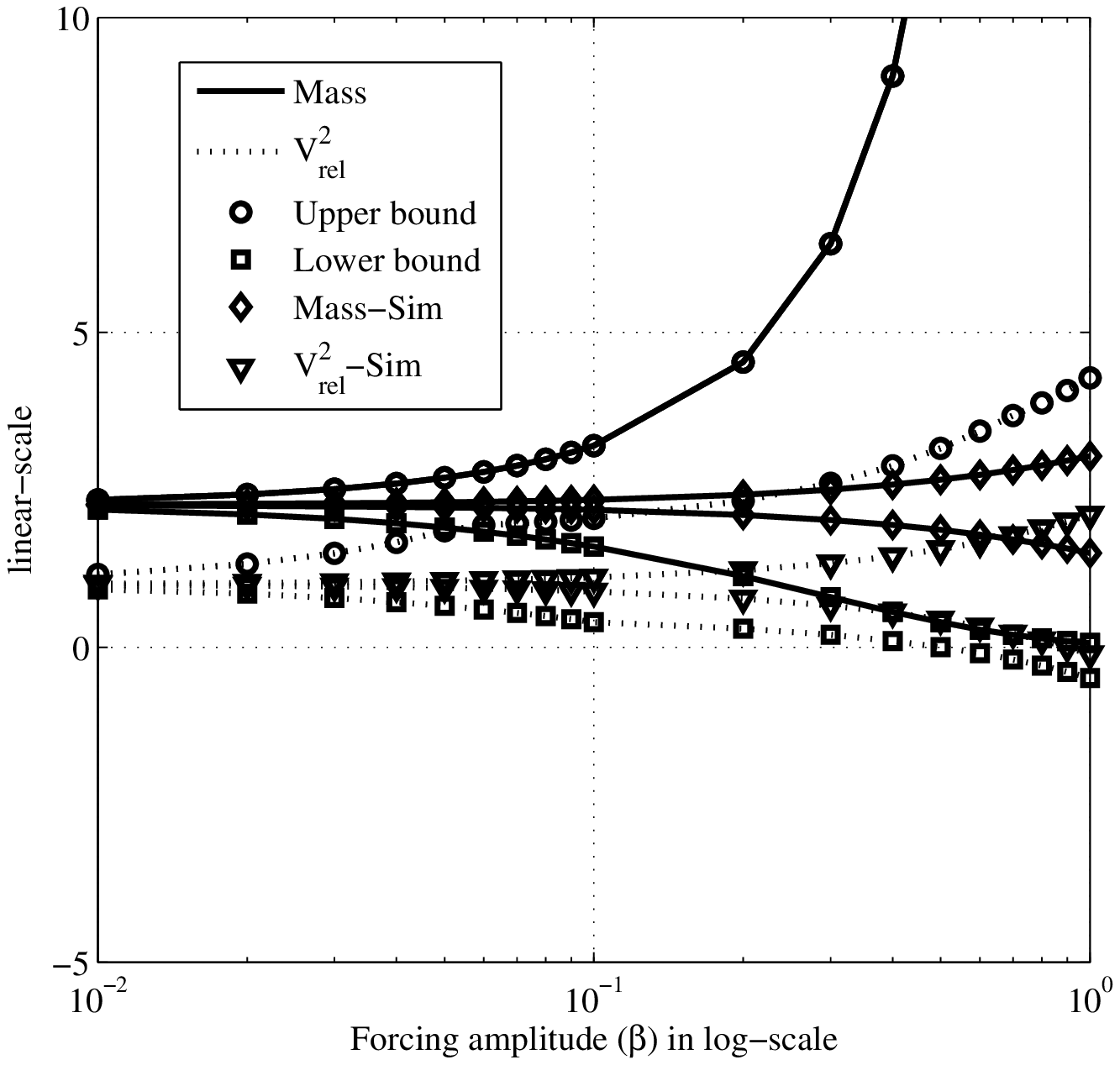}
\par\end{centering}

}
\par\end{centering}

\caption{\textbf{Derived bounds compared to numerical simulations for varying
forcing amplitude $\beta$: }The system parameters corresponds to
the B-ring outer edge where the normal coefficient of restitution
$\varepsilon=0.6$, the forcing period of the moon and the stirring
period are both equal to the orbital period of the ring particles
about Saturn ($T=f_{\mathrm{stir}}=1$). (a) The mean optical depth
is $\tau=1.5$, (b) the mean optical depth is $\tau=1$ and (c) the
mean optical depth is $\tau=0.5$. }
\end{figure}

Next, we study the impact of varying the forcing amplitude $\beta$
on the periodic solutions of $\left(M\left(t\right),v_{\mathrm{rel}}^{2}\left(t\right)\right)$,
and Figures \ref{fig:betaFig1} - \ref{fig:betaFig3} plot the maxima
and minima of the $T$-periodic solution of the $\left(M,v_{\mathrm{rel}}^{2}\right)$
system obtained using numerical simulations against the derived upper
and lower bounds for the $T$-periodic solution of the $\left(M,v_{\mathrm{rel}}^{2}\right)$
system from Theorem \ref{th2} and Theorem \ref{th3}. The system
parameters are the same as before and the stirring period is assumed
to be equal to the orbital period of the ring particles around Saturn,
in other words, $f_{\mathrm{stir}}=1$.

Notice that for small forcing amplitudes, the periodic solutions for
$M\left(t\right)$ and $v_{\mathrm{rel}}^{2}\left(t\right)$ has a
small amplitude about their average values, which in this case, are
the asymptotically stable fixed points (see Remark \ref{rem:fixedPoint})
of the corresponding autonomous system of differential equations.
With increase in $\beta,$ we see an increasing deviations in the
amplitude of $M\left(t\right)$ about the average values, while still
non-negative, since the lower-bound from (\ref{eq:b1}) is non-negative.
Translating to the physical phenomena, the ring particle aggregates
tend to achieve large masses as well as negligibly small masses, with
increase in the forcing amplitude, in each period.

The increase in forcing amplitudes causes a similar effect on $v_{\mathrm{rel}}^{2}\left(t\right)$
as on $M\left(t\right)$. But increasing the forcing amplitude beyond
a certain threshold causes $v_{\mathrm{rel}}^{2}\left(t\right)$ to
take negative values, which is non-physical. This sets a strict upper
limit on the forcing amplitude, which in the case of the above figure,
happens around $\beta=0.6.$ Comparing Figure \ref{fig:betaFig1}
and \ref{fig:betaFig3}, as $\tau$ increases from 0.5 to 1.5, $M\left(t\right)$
maxima increases, $v_{\mathrm{rel}}^{2}\left(t\right)$ maxima changes
only slightly.

Finally, notice that the upper and lower bounds for $M\left(t\right)$
and $v_{\mathrm{rel}}^{2}\left(t\right)$ are tight for small forcing
amplitudes. This means that the behavior of the system is well-characterized
by the bound derived in Theorem \ref{th2} and Theorem \ref{th3}
for the regions in the Saturn's ring system where the influence of
the moons are negligible. The analytical results provide a reasonable
idea about the nature of the system and parallel the explicit numerical
simulations.

\section{Conclusions}

We consider the mathematical model developed in \cite{Es} for understanding
the dynamics of certain specific locations in the Saturn's ring system
due to the periodic perturbations caused by the nearby moon. This
model relates the mean aggregate mass and the velocity dispersion
of the ring particles at a certain location in the Saturn's ring system
in a manner similar to the popular predator-prey model in ecology,
although it represents a totally different physical phenomenon. Further,
this has been studied purely using numerical simulations so far.

In this paper, we conduct a rigorous mathematical analysis of the
dynamical system and derive results pertaining to the existence, uniqueness
and stability of its solution. We analytically prove that the dynamical
system causes a periodic solution for the mean aggregate mass and
the velocity dispersion of the ring particles at a period equal to
the forcing period of the moon. Further, upper and lower bounds for
the solution to the system are derived, that are especially tight
for small perturbations caused by the moons. These bounds well-characterize
the dependence of the system solution on various system parameters.
Several useful insights are drawn about the Saturn's ring system based
on the theoretical results presented here.


\appendix

\section{\label{sub:proofth1}Proof for Theorem \ref{th1}}

Deriving the first equation and inserting the second one, we obtain
the following second order differential equation of Duffing type
\begin{equation}
z''+cz'+bd\exp(2z)=ac+bf(t).\label{duff}
\end{equation}
Now the periodic problems for $(\ref{model2})$ and $(\ref{duff})$
are completely equivalent. If $(x,y)$ is a $T$-periodic solution
of system $(\ref{model2})$ with $x(t)>0$ for all $t$, then $z=\ln x$
is a $T$-periodic solution of $(\ref{duff})$. Conversely, if $z$
is a $T$-periodic solution of eq. $(\ref{duff})$, then $x=\exp(z),y=(a-z')/b$
is a $T$-periodic solution of system $(\ref{model2})$ with $x(t)>0$
for all $t$.

Since $ac>0$, eq. $(\ref{duff})$ is of Landesman-Lazer type. Landesman-Lazer
conditions are classical and implies the existence of a periodic solution,
see for instance \cite[Theorem 1]{Ward}.

\section{\label{sec:proofStab1}Proof for Theorem \ref{stab1}}

We know that system $(\ref{model2})$ is equivalent to eq. $(\ref{duff})$,
hence along this proof we will work directly with this last equation.

{\textbf{Uniqueness.} Assume that $z_{1},z_{2}$ are two $T$-periodic
solutions of eq. $(\ref{duff})$. The difference $\delta(t)=z_{1}(t)-z_{2}(t)$
is a solution of a second order linear equation
\begin{equation}
\delta''+c\delta'+h(t)\delta=0,\label{lin}
\end{equation}
where $h(t)=bd\frac{\exp(2z_{1})-\exp(2z_{2})}{z_{1}-z_{2}}$. We
consider two cases: }
\begin{itemize}
\item \textit{Case 1}: $\delta(t)\ne0$ for every $t$: without loss of
generality, we may take $\delta(t)>0$. By the Mean Value Theorem,
\[
h(t)=2bd\exp(2\xi(t)),
\]
where $z_{2}(t)<\xi(t)<z_{1}(t)$ for every $t$. Hence, by using
$(\ref{mean})$, one obtains
\[
\norm{h}_{1}=2bd\int_{0}^{T}\exp(2\xi(t))dt<2bd\int_{0}^{T}\exp(2z_{1}(t))dt=2acT.
\]
Using $(\ref{cond1})$, one may check easily that $h\in\Omega_{1,c}$,
as defined in \cite[Section 3]{To}. Then, by \cite[Corolllary 2.5]{To},
the operator defined by the left-hand side of $(\ref{lin})$ is inversely
positive, in particular $(\ref{lin})$ is non-degenerate and its only
periodic solution is the trivial one, which is a contradiction with
the assumption that $\delta(t)\ne0$ for every $t$.
\item \textit{Case 2}: $\delta(t_{0})=0$ for some $t_{0}\in[0,T]$: by
periodicity, there exists $t_{1}>t_{0}$ such that $d(t_{1})=0$ and
$\delta(t)\ne0$ for all $t\in]t_{0},t_{1}[$. Again, we may assume
without losing generality that $\delta(t)>0$ for all $t\in]t_{0},t_{1}[$.
Define the truncated function $\hat{h}(t)$ as $h(t)$ if $t\in[t_{0},t_{1}]$
and $0$ if $t\in[0,t_{0}[\cup]t_{1},T]$. By reasoning as before,
$h\in\Omega_{1,c}$. Then, as a consequence of the results in \cite{To}
(see Remark 2.2 therein), the distance between two consecutive zeroes
of a solution of $\delta''+c\delta'+\hat{h}(t)\delta=0$ is greater
than $T$, but $\delta$ is a solution vanishing at $t_{0},t_{1}$,
which is a contradiction.
\end{itemize}
After this analysis, the only remaining possibility is $d(t)=0$ for
every $t$, so the $T$-periodic solution is unique.

{\textbf{Stability.} The proof for asymptotic stability is also based
on the results from \cite{To}. Let $z(t)$ be the unique $T$-periodic
solution of eq. $(\ref{duff})$. A standard computation gives that
the topological index of $z$ is $\gamma(z)=1$. The variational equation
is $u''+cu'+h(t)u=0$ with $h(t)=2bd\exp(2z(t))$ and again $h\in\Omega_{1,c}$
as a result of $(\ref{cond1})$. Then the asymptotic stability follows
directly from \cite[Theorem 3.1]{To}.}

\section{\label{sec:proofStab2}Proof for Theorem \ref{stab2}}

We use the terminology and notation from \cite{NO}. A lower solution
is found by fixing a constant $\alpha$ such that
\begin{equation}
bd\exp(2\alpha)=ac+b\norm{f}_{\infty}.\label{alpha}
\end{equation}
An upper solution is easily obtained as follows. Take $F$ the unique
$T$-periodic solution of $z''+cz'=f(t)$ with mean value zero. Then,
$\beta(t)=-M+F(t)$ with $M$ big enough is an upper solution with
$\beta<\alpha$ . To apply \cite[Theorem 1.2]{NO}, we have to check
two hypotheses. The fist one is the finiteness of the number of $T$-periodic
solutions, which is direct from the analiticity of the field on the
state variables. The second hypothesis is condition (1.6) in \cite{NO}.
Fix $g(t,z)=bd\exp(2z)-ac-bf(t)$. For $s\in[\beta(t),\alpha]$,
\[
\frac{\partial}{\partial s}g(t,s)=2bd\exp(2s)\leq2bd\exp(2\alpha)=2ac+2b\norm{f}_{\infty}<\frac{c^{2}}{4}+\frac{\pi^{2}}{T^{2}}
\]
by $(\ref{cond2})$, which is just (1.6) in \cite{NO}. The result
is done.

\section{\label{sec:proofth2}Proof for Theorem \ref{th2}}

We work again over the equivalent eq. $(\ref{duff})$. Integrating
$(\ref{duff})$ over a whole period
\begin{equation}
bd\int_{0}^{T}\exp(2z)dt=acT.\label{mean}
\end{equation}
By the integral mean value theorem, there exists $t_{0}\in[0,T]$
such that
\begin{equation}
\exp(2z(t_{0}))=\frac{ac}{bd}.\label{t0}
\end{equation}
Observe that $z(t)-z(t_{0})\in H_{0}^{1}(t_{0},T+t_{0})$, then by
the Sobolev inequality with the optimal constant (see for instance
\cite{Ta}), we get
\begin{equation}
\abs{z(t)-z(t_{0})}\leq\frac{\sqrt{T}}{2}\norm{z'}_{2}.\label{ineq1}
\end{equation}
On the other hand, multiplying $(\ref{duff})$ by $z'$ and integrating
over $[0,T]$ and applying Cauchy-Schwarz inequality, we get
\[
c\norm{z'}_{2}^{2}=b\int_{0}^{T}f(t)z'dt\leq b\norm{f}_{2}\norm{z'}_{2},
\]
hence,
\begin{equation}
\norm{z'}_{2}\leq\frac{b}{c}\norm{f}_{2}.\label{ineq2}
\end{equation}
Now, combining $(\ref{t0})$,$(\ref{ineq1})$ and $(\ref{ineq2})$,
\[
z(t)=z(t_{0})+z(t)-z(t_{0})\leq\frac{1}{2}\ln\left(\frac{ac}{bd}\right)+\frac{\sqrt{T}}{2}\frac{b}{c}\norm{f}_{2}
\]
and
\[
z(t)=z(t_{0})+z(t)-z(t_{0})\geq\frac{1}{2}\ln\left(\frac{ac}{bd}\right)-\frac{\sqrt{T}}{2}\frac{b}{c}\norm{f}_{2}.
\]
The first inequality of $(\ref{bounds})$ arises from here taking
$z=\ln x$ and finding the value of $x$.

The bound for $y$ is found as follows. Let $t_{*}$ such that $z(t_{*})=\min_{t\in[0,T]}z(t)$.
Integrating $(\ref{duff})$ on $[t_{*},t]$,
\[
z'(t)+cz(t)-cz(t_{*})+bd\int_{t_{*}}^{t}\exp(2z)dt=\int_{t_{*}}^{t}ac+bf(s)ds\leq acT+b\norm{f^{+}}_{1}.
\]
Since $cz(t)-cz(t_{*})+bd\int_{t_{*}}^{t}\exp(z)dt\geq0$, then
\begin{equation}
z'(t)\leq\int_{t_{*}}^{t}ac+bf(s)ds\leq acT+b\norm{f^{+}}_{1}.\label{der1}
\end{equation}
Analogously, Let $t^{*}$ such that $z(t^{*})=\max_{t\in[0,T]}z(t)$.
Integrating $(\ref{duff})$ on $[t^{*},t]$,
\[
z'(t)+cz(t)-cz(t^{*})+bd\int_{t^{*}}^{t}\exp(2z)dt=\int_{t^{*}}^{t}ac+bf(s)ds\geq-b\norm{f^{-}}_{1}.
\]
Then,
\[
z'(t)\geq-bd\int_{t^{*}}^{t}\exp(z)dt-b\norm{f^{-}}_{1}\geq-bd\int_{t^{*}}^{t^{*}+T}\exp(z)dt-b\norm{f^{-}}_{1}.
\]
Since $z(t)$ is $T$-periodic, $\int_{t^{*}}^{t^{*}+T}\exp(2z)dt=\int_{0}^{T}\exp(2z)dt$,
then by using $(\ref{mean})$,
\begin{equation}
z'(t)\geq-acT-b\norm{f^{-}}_{1}.\label{der2}
\end{equation}
Finally, inserting $(\ref{der1})-(\ref{der2})$ into $y=\frac{1}{b}(a-z')$,
one obtains the second inequality of $(\ref{bounds})$.

\section{\label{sec:proofth3}Proof for Theorem \ref{th3}}

Note that
\[
\frac{1}{\exp(cT)-1}\leq G(t,s)\leq\frac{\exp(cT)}{\exp(cT)-1}
\]
for all $t,s\in[0,T]$.

In the following steps, we derive upper and lower bounds for both
the terms in the above expression, which will be used later to obtain
bounds for $y\left(t\right).$

Bounds for $u\left(t\right)=\int_{s=t}^{t+T}G\left(t,s\right)\exp\left(2z\left(s\right)\right)ds$:

Using the bounds for $G\left(t,s\right)$,
\begin{eqnarray}
\frac{acT}{bd\left(\mathrm{e}^{cT}-1\right)} & \le u\left(t\right)\le & \frac{acT\mathrm{e}^{cT}}{bd\left(\mathrm{e}^{cT}-1\right)}.\label{eq:ufunbounds1}
\end{eqnarray}
The next set of bounds are derived by using the bounds derived for
$z\left(t\right)$ in the proof of Theorem \ref{th2} and noting that
$\int_{s=t}^{t+T}G\left(t,s\right)ds=\frac{1}{c}$
\begin{eqnarray}
\frac{a}{bd}\exp\left(-\frac{b\sqrt{T}\left\Vert f\right\Vert _{2}}{c}\right) & \le u\left(t\right)\le & \frac{a}{bd}\exp\left(\frac{b\sqrt{T}\left\Vert f\right\Vert _{2}}{c}\right).\label{eq:ufunbounds2}
\end{eqnarray}
Combining (\ref{eq:ufunbounds1}) and (\ref{eq:ufunbounds2}) , we
get
\begin{equation}
\max\left(\frac{acT}{bd\left(\mathrm{e}^{cT}-1\right)},\frac{a}{bd}\exp\left(-\frac{b\sqrt{T}\left\Vert f\right\Vert _{2}}{c}\right)\right)\le u\left(t\right)\le\min\left(\frac{acT\mathrm{e}^{cT}}{bd\left(\mathrm{e}^{cT}-1\right)},\frac{a}{bd}\exp\left(\frac{b\sqrt{T}\left\Vert f\right\Vert _{2}}{c}\right)\right).\label{eq:ufunbound}
\end{equation}

Bounds for $v\left(t\right)=\int_{s=t}^{t+T}G\left(t,s\right)f\left(s\right)ds$:

Let $p\left(t,s\right)\triangleq\frac{G\left(t,s\right)}{\int_{s=t}^{t+T}G\left(t,s\right)ds}.$
We know, $\int_{s=t}^{t+T}G\left(t,s\right)ds=\frac{1}{c}.$ Hence,
$v\left(t\right)=\frac{1}{c}\int_{s=t}^{t+T}p\left(t,s\right)f\left(s\right)ds.$
Since $f_{min}\le\int_{s=t}^{t+T}p\left(t,s\right)f\left(s\right)ds\le f_{max},$
we have the first set of bounds
\begin{eqnarray}
\frac{f_{min}}{c} & \le v\left(t\right)\le & \frac{f_{max}}{c}.\label{eq:vfunbounds1}
\end{eqnarray}
Similarly, using the Cauchy-Schwartz inequality, we have $\int_{s=t}^{t+T}p\left(t,s\right)f\left(s\right)ds\le\left\Vert f\right\Vert _{2}\sqrt{\int_{s=t}^{t+T}p^{2}\left(t,s\right)ds}$
and hence
\begin{eqnarray}
v\left(t\right) & \le & \left\Vert f\right\Vert _{2}\sqrt{\frac{\mathrm{e}^{cT}+1}{2c\left(\mathrm{e}^{cT}-1\right)}}.\label{eq:vfunbounds2}
\end{eqnarray}

Another set of bounds based on the following inequality
\begin{eqnarray*}
-\int_{s=t}^{t+T}G\left(t,s\right)f^{-}\left(s\right)ds & \le v\left(t\right)\le & \int_{s=t}^{t+T}G\left(t,s\right)f^{+}\left(s\right)ds
\end{eqnarray*}
are
\begin{eqnarray}
\frac{-\left\Vert f^{-}\right\Vert _{\infty}}{c} & \le v\left(t\right)\le & \frac{\left\Vert f^{+}\right\Vert _{\infty}}{c}\label{eq:vfunbounds3}\\
\frac{-\left\Vert f^{-}\right\Vert _{1}\mathrm{e}^{cT}}{\left(\mathrm{e}^{cT}-1\right)} & \le v\left(t\right)\le & \frac{\left\Vert f^{+}\right\Vert _{1}\mathrm{e}^{cT}}{\left(\mathrm{e}^{cT}-1\right)}.\label{eq:vfunbounds4}
\end{eqnarray}
And finally, if the $v\left(t\right)$ is computable for all $t$,
then,
\begin{eqnarray}
v_{min} & \le v\left(t\right)\le & v_{max}.\label{eq:vfunbounds5}
\end{eqnarray}
Note that (\ref{eq:vfunbounds1}) and (\ref{eq:vfunbounds3}) are
equivalent. Hence, combining (\ref{eq:vfunbounds2}) - (\ref{eq:vfunbounds5}),
we get
\begin{eqnarray}
v\left(t\right) & \ge & \min\left(v_{min},\frac{-\left\Vert f^{-}\right\Vert _{\infty}}{c},\frac{-\left\Vert f^{-}\right\Vert _{1}\mathrm{e}^{cT}}{\left(\mathrm{e}^{cT}-1\right)}\right)\nonumber \\
v\left(t\right) & \le & \min\left(v_{max},\left\Vert f\right\Vert _{2}\sqrt{\frac{\mathrm{e}^{cT}+1}{2c\left(\mathrm{e}^{cT}-1\right)}},\frac{\left\Vert f^{+}\right\Vert _{\infty}}{c},\frac{\left\Vert f^{+}\right\Vert _{1}\mathrm{e}^{cT}}{\left(\mathrm{e}^{cT}-1\right)}\right).\label{eq:vfunbounds}
\end{eqnarray}
Using (\ref{eq:ufunbound}) and (\ref{eq:vfunbounds}), (\ref{eq:ylb})
and (\ref{eq:yub}) follow directly.
\end{document}